\documentclass[a4paper,11pt]{article}
\pdfoutput=1 
\usepackage{jcappub}

\usepackage{graphicx}
\usepackage{longtable}
\usepackage{float}
\usepackage{dcolumn}
\usepackage{bm}
\usepackage{appendix}
\usepackage{multirow}
\usepackage{color}
\usepackage[utf8]{inputenc}
\usepackage{footmisc}
\newcommand{\mytilde}{\raise.17ex\hbox{$\scriptstyle\mathtt{\sim}$}}
\newcommand{\barr}{\begin{eqnarray}}
\newcommand{\earr}{\end{eqnarray}}
\newcommand{\bea}{\begin{eqnarray*}}
\newcommand{\eea}{\end{eqnarray*}}
\newcommand{\beq}{\begin{equation}}
\newcommand{\eeq}{\end{equation}}

\newcommand{\Mpc}{\mathrm{Mpc}}

\renewcommand{\bf}{\rm}
\title{Looking for interactions in the cosmological dark sector}

\author[a, b,c]{M. Benetti} \emailAdd{benettim@na.infn.it}

\author[d]{W. Miranda} \emailAdd{mirandawelber@gmail.com}

\author[e]{H. A. Borges}\emailAdd{humberto@ufba.br}

\author[e]{C. Pigozzo} \emailAdd{cpigozzo@ufba.br}

\author[e]{S. Carneiro} \emailAdd{saulo.carneiro.ufba@gmail.com}

\author[c]{J. S. Alcaniz}\emailAdd{alcaniz@on.br}

\affiliation[a]{Dipartimento di Fisica  ``E. Pancini", Universit\`a di Napoli  ``Federico II", Via Cinthia, I-80126, Napoli, Italy}

\affiliation[b]{Istituto Nazionale di Fisica Nucleare (INFN), sez. di Napoli, Via Cinthia 9, I-80126 Napoli, Italy}

\affiliation[c]{Departamento de Astronomia, Observat\'orio Nacional, Rio de Janeiro, RJ, Brasil}

\affiliation[d]{Instituto Federal da Bahia, Paulo Afonso, BA, Brasil}

\affiliation[e]{Instituto de F\'{\i}sica, Universidade Federal da Bahia, Salvador, BA, Brasil}

\abstract{
We study observational signatures of non-gravitational interactions between the dark components of the cosmic fluid, which can be either due to creation of dark particles from the expanding vacuum or an effect of the clustering of a dynamical dark energy. In particular, we analyse a class of interacting models ($\Lambda$(t)CDM), characterised by the parameter $\alpha$, that behaves at background level like cold matter at early times and tends to a cosmological constant in the asymptotic future. In our analysis we consider both background and primordial perturbations evolutions of the model. We use Cosmic Microwave Background (CMB) data together with late time observations, such as the Joint Light-curve Analysis (JLA) supernovae data, the Hubble Space Telescope (HST) measurement of the local value of the Hubble-{Lema\^itre} parameter, and primordial deuterium abundance from Ly$\alpha$ systems to test the observational viability of the model and some of its extensions. We found that there is no preference for values of $\alpha$ different from zero (characterising interaction), even if there are some indications for positive values when the minimal $\Lambda$(t)CDM model is analysed.  When extra degrees of freedom in the relativistic component of the cosmic fluid are considered, the data favour negative values of $\alpha$, which means an energy flux from dark energy to dark matter.}

\begin{document}
\maketitle

\section{Introduction}
\label{Introduction}

The standard cosmological model, also known as $\Lambda$ - Cold Dark Matter ($\Lambda$CDM), provides a successful description of the structure and evolution of the universe, requiring half a dozen parameters. However, in spite of its observational successes, many open questions remain, including the very nature of the dark matter and dark energy components, which drive the current cosmic evolution. Furthermore, as the accuracy of cosmological observations increases -- some of the current constraints on the $\Lambda$CDM parameters can reach sub-percent level -- tensions between different data sets have also emerged. 

This is the case for instance of the discrepancy involving current measurements of the Hubble-{Lema\^itre} parameter, in which the value obtained from CMB data assuming the $\Lambda$CDM model, $H_0 = 66.93 \pm 0.62$ $\rm{km/s/Mpc}$ \cite{Aghanim:2018eyx}, differs by $\simeq 4.4 \sigma$ from the value measured using distance measurements of galaxies in the local Universe calibrated by Cepheid variables and type Ia Supernovae (SNe Ia), $H_0=74.03 \pm 1.42$ $\rm{km/s/Mpc}$ \cite{Riess:2019cxk} (see e.g. \cite{Verde:2019ivm} for a recent discussion). Another ongoing issue concerns the preference of the Planck CMB angular spectra for a large amplitude of the lensing signal, $ \mathcal A_l$. The current CMB data furnish $ \mathcal A_l = 1.18 \pm 0.14$ at 95\% C. L. \cite{Aghanim:2018eyx}, which is about $3\sigma$ off from the $\Lambda$CDM value $ \mathcal A_l = 1 $. This discrepancy is particularly challenging given that the lensing signal obtained from the Planck angular trispectrum is consistent with the standard cosmology \cite{Aghanim:2018eyx}. Furthermore, a $\simeq 2\sigma$ difference in the $\Omega_m - \sigma_8$ plane can be inferred by comparing CMB and cosmic shear data~\cite{Hildebrandt:2018yau} (see also \cite{MacCrann:2014wfa}), where $\Omega_m$ is the matter density parameter and $\sigma_8$ is the matter fluctuation amplitude on scales of $8h^{-1}$Mpc. Currently, one of the main arguments is that such discrepancies may be an indication of new physics beyond $\Lambda$CDM, and many analyses have investigated alternative scenarios that could reconcile the data sets. Attempts to solve the $H_0$-tension problem for instance include extensions of the standard model, such as the existence of new relativistic particles~\cite{pedro,DEramo:2018vss}, early dark energy \cite{Poulin:2018cxd}, primordial gravitational waves \cite{Graef:2018fzu}, small spatial curvature~\cite{Bolejko:2017fos}, among others (see also \cite{Mortsell:2018mfj,Benetti:2017juy} and references therein). In general, it is not straightforward to reconcile both the $H_0$ and $\sigma_8$ discrepancies in the same theoretical framework (see e.g. \cite{Benetti:2017juy}).

On the other hand, there has been a growing interest in models with interaction between the dark components \cite{Billyard:2000bh,Amendola:1999er,Zimdahl:2001ar,He:2008tn,Valiviita:2008iv,Odderskov:2015fba,wiliam,Gonzalez:2018rop,Cid:2018ugy, vonMarttens:2018iav, Dam:2019prv}. A basic hypothesis of standard cosmology is that pressureless matter is conserved, which is mathematically expressed by the well-known evolution law $\rho_m \propto a^{-3}$, where $\rho_m$ is the matter density and $a$ the cosmological scale factor. This hypothesis is in fact a tautology if dark energy is formed by a cosmological constant. However, if dark energy is a dynamical field, it can in principle cluster. In this case, it is impossible to know which part of the clustering energy is made by matter and which part is made by clustered dark energy.  This is a manifestation of the so called {\it dark degeneracy} \cite{Kunz:2007rk, Wasserman:2002gb, Rubano:2002sx,CB}. If we identify the clustered energy as pressureless matter, the clustering of the dynamical dark energy will be interpreted as matter production.  At the same time, the smooth part of dark energy, responsible for accelerating the expansion, will decrease. This is a scenario in which an interaction between dark matter (defined as clustering energy) and dark energy (defined as the smooth component responsible for the acceleration) would appear \cite{1Models,2Models,3Models,jailson,Carneiro:2017iww, Carneiro:2018zet}. Another situation is that dark particles are indeed created from the expanding vacuum, a process generally present in expanding spacetimes. Since we know very little about the properties of dark particles (as their masses and couplings), their creation from vacuum cannot, a priori, be ruled out.

From the observational viewpoint, analyses of these models have shown that the current matter density derived from Large-Scale Structure (LSS) observations are systematically lower than the values obtained through SNe Ia data. The $\Lambda$CDM best-fit for the 2dFGRS data, for instance, is $\Omega_{m0} \approx 0.23$ \cite{2dF}, while a value $\Omega_{m0} \approx 0.24$ was obtained with the linear-range data of the SDSS galaxy catalogue \cite{tegmark,sdss}\footnote{When non-linear scales are included, one finds $\Omega_{m0} \approx 0.29$ \cite{sdss, tegmark2}.}. On the other hand, from type Ia supernovae (SNe Ia) samples larger densities are obtained. For example, the JLA compilation furnishes $\Omega_{m0} \approx 0.3$ \cite{JLA} or even $\Omega_{m0} \approx 0.4$, depending on the light-curve calibration method used \cite{trotta}. Larger values are also derived from CMB observations \cite{Aghanim:2018eyx}. This difference may be understood as a signature of matter creation since the matter power spectrum depends sensibly on the matter density at the time of matter-radiation equality, whereas cosmological distances determinations are more dependent on the present matter density. Therefore, if matter is created in the late-time expansion, but we assume matter conservation in LSS and distance tests, the former will lead to a lower present density as compared to the latter. The $H_0$-tension mentioned earlier has also been discussed in the context of interacting models of dark energy (see e.g. \cite{Pandey:2019plg, Pan:2019gop, DiValentino:2019ffd}).  

Here, we study a class of interacting model that, at background level, can be associated with a Generalised Chaplygin Gas (GCG) model. The observational viability of the model is analysed at both background and  perturbative levels using the current data of SNe Ia and CMB, along with measurements of $H_0$ and primordial deuterium abundance from Ly$\alpha$ systems\footnote{Soon after the completion of the analyses presented in this work the latest CMB data from the Planck Collaboration was made available.}. We organise this paper as follows. In Section \ref{Sec:Theory} we introduce the theory of the model, discussing the background evolution as well as the primordial perturbations formula. Details of the data used and the analysis performed are given in Section \ref{Sec:Data} and Section \ref{Sec:Analysis} along with the results of our statistical analysis. Finally, discussions and conclusions are presented in Section \ref{Sec:Conclusions}.

\section{Parametrising the interactions}
\label{Sec:Theory}
In a FLRW spacetime the energy-momentum tensor has the general form
\begin{equation}
T^{\mu \nu} = (\rho + p) u^{\mu} u^{\nu} - p g^{\mu \nu},
\end{equation}
where $\rho$ is the energy density, $p = \omega(\rho) \rho$ is the pressure, and $u$ is the cosmic fluid $4$-velocity. One may decompose this perfect fluid as
\begin{equation}
T^{\mu \nu} = \rho_m u^{\mu} u^{\nu} + \Lambda g^{\mu \nu},
\end{equation}
by defining $\Lambda = -p = - \omega \rho$ and $\rho_m = (1+ \omega) \rho$. Such a decomposition resolves the degeneracy discussed above, provided we can show that the vacuum-type component $\Lambda$ does not cluster if we identify $\rho_m$ as the observed cold matter. For this purpose, let us express the covariant conservation equation $T^{\mu \nu}_{;\nu} = 0$ in the form
\begin{eqnarray} \label{conservation1}
T^{\mu \nu}_{m;\nu} &=& Q^{\mu},\\ \label{conservation2}
T^{\mu \nu}_{\Lambda;\nu} &=& - Q^{\mu},
\end{eqnarray}
where $T^{\mu \nu}_m = \rho_m u^{\mu} u^{\nu}$ and $T^{\mu \nu}_{\Lambda} = \Lambda g^{\mu \nu}$. Here, $Q^{\mu}$ is the energy-momentum transfer between the two components, which we decompose as $Q^{\mu} = Q u^{\mu} + \bar{Q}^{\mu}$, with $\bar{Q}^{\mu} u_{\mu} = 0$. For a comoving observer, it is easy to see that $Q$ represents the energy transfer. The second term, $\bar{Q}^{\mu}$, represents the momentum transfer, null in the isotropic background. From (\ref{conservation2}) we have
\begin{eqnarray}
Q &=& - \Lambda_{,\nu} u^{\nu},\\
\bar{Q}^{\mu} &=& \Lambda_{,\nu} (u^{\mu} u^{\nu} - g^{\mu \nu}).
\end{eqnarray}
A linear perturbation of the above equations leads to $\delta \bar{Q}^0 = 0$ and
\begin{eqnarray}\label{Q}
\delta Q &=& - (\delta \Lambda)_{,0} + \frac{\dot{\Lambda}}{2}\,\delta g_ {00},\\
\delta \bar{Q}_i &=& (\delta \Lambda + \dot{\Lambda} \theta)_{,i} \equiv \delta \Lambda^c_{,i},
\end{eqnarray}
where $\theta$ is the dark fluid velocity potential. The last equation plays an important role in our analysis~\cite{CB}. The left-hand side represents the momentum transfer between the two components in the perturbed spacetime. If the matter component is non-relativistic, the momentum transfer must be negligible, and matter will follow geodesics \cite{Koyama:2009gd, Wands:2012vg, winfried}. We then obtain $ \delta \Lambda^c \approx 0$. This quantity represents the covariant, comoving perturbation of the $\Lambda$ component \cite{Malik}, which is, therefore, smooth. From (\ref{Q}) we also see that $\delta Q \approx 0$ in a synchronous comoving gauge. If, in addition, there is no energy transfer between the two components, $\Lambda$ is constant and matter is conserved\footnote{The reverse is also true: If we assume that clustering cold matter is conserved, observations will favor a cosmological constant among dark energy candidates.}. In this case, observational cosmology would be reduced to a precise determination of the $\Lambda$CDM free parameters. However, there is no reason to assume such a prior, which demands for observational confirmation.

\subsection{Background}
With the above decomposition of the cosmic fluid, the Friedmann and conservation equations assume the form
\begin{eqnarray} \label{Friedmann}
3H^2 = \rho_m + \Lambda,\\ \label{conservation}
\dot{\rho}_m + 3H\rho_m = \Gamma \rho_m = -\dot{\Lambda},
\end{eqnarray}
where the unknown function $\Gamma$ is the rate of matter creation. Note that the second equality is quite general and can be seen as a definition of the creation rate. In the late-time universe there are two natural possibilities for fixing $\Gamma$. The first is a constant creation rate \cite{PLB}, a natural choice for the creation of non-relativistic particles in a low-energy expansion. The second is a creation rate proportional to the expansion rate $H$. The former (latter) is a particular (limiting) case of a parametrisation given by the ansatz
\begin{equation} \label{Lambda}
\Lambda = \sigma H^{-2\alpha},
\end{equation}
with constants $\alpha > -1$ and $\sigma = 3 (1 - \Omega_{m0}) H_0^{2(\alpha+1)}$. From (\ref{Friedmann}) and (\ref{conservation}) it is easy to derive
\begin{equation}
\Gamma = -\alpha \sigma H^{-(2\alpha +1)},
\end{equation}
and, including a conserved radiation component, we obtain the Hubble function of our model, that hereafter we call $\Lambda$(t)CDM model, as
\begin{equation}\label{eq:E}
E(z) = {H(z)}/{H_0} = \sqrt{\left[ (1-\Omega_{m0}) + \Omega_{m0} (1+z)^{3(1+\alpha)} \right]^{\frac{1}{(1+\alpha)}} + \Omega_{R0} (1+z)^4}.
\end{equation}
 The $\Lambda$CDM model corresponds to $\alpha = 0$. Negative values of $\alpha$ means creation of matter, while for $\alpha = -1/2$ we have a constant creation rate $\Gamma = 3H_{dS}/2$, where $H_{dS}^{-1}$ is the asymptotic future de Sitter horizon. The reader may identify Eq. (\ref{eq:E}) with a generalised Chaplygin gas (GCG) \cite{cg1,Alcaniz:2002yt,cg2,cg3,cg4}, that behaves like cold matter at early times and tends to a cosmological constant in the asymptotic future. This identification, however, is only valid at the background level. Our ansatz (\ref{Lambda}) is, actually, equivalent to a decomposed, non-adiabatic GCG \cite{non-adiabatic,non-adiabatic2,wands2,bb}. This parametrisation, of course, does not include all the possible forms for the interaction term. Nevertheless, it is general enough for our purpose, namely to look for signatures of interactions in the current observational data. From Eq. (\ref{eq:E}) it is easy to see that, for high redshifts, the matter density scales as
\begin{equation} \label{density_high}
\rho_m(z) = 3 H_0^2\, \Omega_{m0}^{\frac{1}{1+\alpha}} z^3 \quad (z \gg 1).
\end{equation}
Therefore, owing to matter creation, we do not have the standard relation between the matter density at high redshifts and at present. This difference affects some expressions used in fitting formulae and numerical codes, as for example equation (\ref{k}) below. Furthermore, in order to have the correct density at high redshifts (preserving in this way the CMB spectrum profile), we will have today a density higher or smaller than the standard model value (depending on the sign of $\alpha$).

\subsubsection{Implication in the matter power spectrum}
\begin{figure}
\includegraphics[height=5cm]{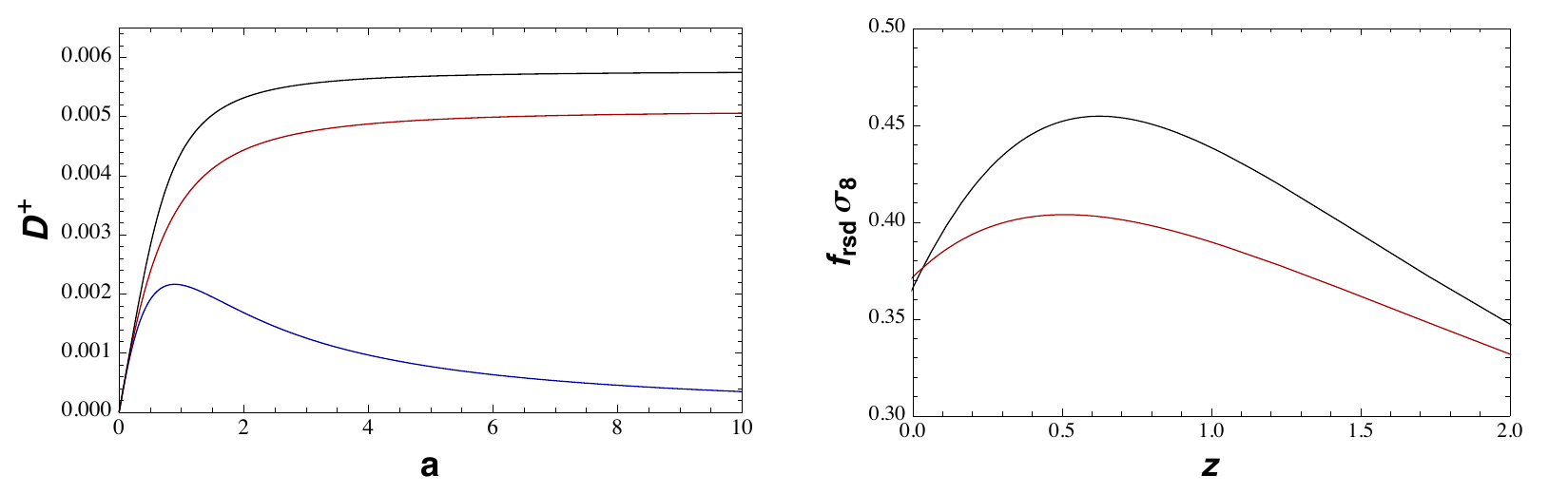}
\caption{{\bf Left panel:} Evolution of the matter contrast in the $\Lambda$CDM model (black curve), and of the total matter (blue) and baryonic (red) contrasts for  $\alpha = -1/2$ \cite{Velten}. {\bf Right panel}: Evolution of the baryonic $f_{rsd} \sigma_8$ for $\Lambda$CDM (black) and for $\alpha = -1/2$ (red) \cite{CB2}.}
\label{fig:contrast}
\end{figure}
Since the only clustering component is pressureless matter, for sub-horizon scales there is no scale-dependence in the perturbation equations governing the evolution of the matter linear contrast. Therefore, all the observed modes evolve in the same way, and the matter power spectrum retains at present the profile it has at the beginning of the matter era. It can be written as \cite{BBKS}
\begin{equation} \label{BBKS}
P(k) = P_0 k^{n_s} T^2(k/k_{eq}),
\end{equation}
where $T(x \equiv k/k_{eq})$ is 
the transfer function.
The normalisation constant $P_0$ is determined in the fitting and can be related to $\sigma_8$ by \cite{Devi}
\begin{equation}
\sigma_8^2 = \frac{1}{2 \pi^2} \int^{\infty}_0 k^2 W^2(kR) P(k) dk,
\end{equation}
with a spherical top-hat filter defined as
\begin{equation}
W(kR) = 3 \left[ \frac{\sin(kR)}{(kR)^3} - \frac{\cos(kR)}{(kR)^2} \right],
\end{equation}
and $R = 8$ Mpc. The spectrum turnover occurs at $k_{eq}$, the mode which enters in the horizon at the time of matter-radiation equality. By using (\ref{density_high}) for the matter density at high redshifts and $\rho_R \approx \rho_{R0} z^4$ for the radiation density, we obtain
\begin{equation} \label{k}
k_{eq} = \frac{H(z_{eq})}{z_{eq}} = 0.073\, \text{Mpc}^{-1} h^2\, \Omega_{m0}^{\frac{1}{1+\alpha}},
\end{equation}
where $z_{eq}$ is the redshift of matter-radiation equality, and $h = H_0/$(100km/s/Mpc). For $\alpha = 0$ we recover the standard expression for $k_{eq}$ \cite{Dodelson}. 

{  Until now we referred to the total matter density $\Omega_{m0}$ but, in order to study the growing function, that is, the evolution of the observed spectrum with time, baryons must be explicitly considered}. The linear contrast of dark matter suffers a late-time suppression owing to dark matter creation. This does not affect the spectrum profile, which, as discussed above, is preserved during the matter era. It only affects its present normalisation, which is correctly fixed by the data. However, it does lead to a suppression in the dark matter growing rate as compared to the standard model \cite{Borges&Wands}. Nevertheless, we should have in mind that observations of the linear power spectrum and redshift space distortions (RSD) refer to visible matter and, as baryons are conserved, there is no late-time suppression in their linear contrast\footnote{One may argue that weak lensing and CMB are sensitive to both baryonic and dark matter distributions. Actually, these observations are sensitive to the gravitational potential generated by total matter. Despite the suppression in the dark matter contrast, the gravitational potential is {  less affected} \cite{non-adiabatic,Velten}, because it is proportional to $\delta \rho_m$, not to $\delta_m$.}.

As an example, let us take again the particular case $\alpha = -1/2$. The baryonic contrast $\delta_b$ can be obtained by a numerical integration of the perturbation equations\footnote{The constant $\Gamma$ in (\ref{humberto}) is the creation rate of total matter, not dark matter. The resulting difference in the background solution is negligible, because baryons are sub-dominant \cite{Velten}.} \cite{Velten}
\begin{eqnarray} \label{humberto}
a^2H^2 \delta_m''+ aH \left( a H' + 3H + \Gamma \right) \delta_m'+ 2 \Gamma H \delta_m &=& \frac{\rho_m \delta_m}{2},\\
\label{humberto_b}
a^2H^2 \delta_b''+ aH \left( a H' + 3H \right) \delta_b' &=& \frac{\rho_m \delta_m}{2},
\end{eqnarray}
where a prime means derivative w.r.t. the scale factor $a$, and the index $m$ stands for total matter. It is straightforward to generalise them to any $\alpha$. The corresponding growing rate is given by
\begin{equation}
f_b(z) = \frac{\delta_b^{'}(z)}{Ha \delta_b(z)},
\end{equation}
and RSD can be tested by using the bias-independent combination
\begin{equation}
f_{rsd}(z) \sigma_8(z) = f_b(z) \sigma_8 D_{b+}^N(z),
\end{equation}
where $D_{b+}^N(z) = \delta_b(z)/\delta_b(0)$ and $\sigma_8 = \sigma_8(0) \approx 0.83$ \cite{Ade:2015zua}. The results are shown in Fig. \ref{fig:contrast}. In the left panel we can see the evolution of total matter contrast in both standard and interacting models, together with the baryonic contrast evolution in the latter, with same initial conditions \cite{Velten}. In the right panel we plot $f_{rsd} \sigma_8$ for the standard and interacting models, which differ at most 12\% \cite{CB2}. For the interacting model we have used $\Omega_{m0} = 0.45$, while for the standard model we have taken $\Omega_{m0} = 0.23$, the best-fit values for the 2dFGRS data in each case \cite{PLB,2dF}.

\subsection{Primordial perturbations}

The Boltzmann equations for conserved baryons and radiation are the same as in the $\Lambda$CDM model. For the $\Lambda$(t)CDM model analysed here, assuming that there is no momentum transfer in the dark matter rest frame, the Poisson and dark matter perturbation equations can be obtained in the longitudinal gauge as\footnote{We adopt here the signature ($-$,$+$,$+$,$+$), with $\phi = \psi$.},
\begin{equation}\label{thetad2}
\theta'_{dm}+\mathcal{H}\theta_{dm}-k^2\Phi=0,
\end{equation}
\begin{equation}\label{deltad2}
\delta'_{dm}-3\Phi'+\theta_{dm}=-\frac{aQ}{\rho_{dm}} \left[ \delta_{dm} - \frac{1}{k^2} \left( k^2 \Phi + \frac{Q'}{Q}\theta_{dm} \right) \right],
\end{equation}
\begin{equation}
-k^2 \Phi=\frac{a^2}{2} (\rho_{dm} \delta_{dm} + \rho_b \delta_b) - \left( \frac{a^3 Q}{2} - \frac{3a^2}{2}\mathcal{H}\rho_m \right)\frac{\theta_{dm}}{k^2},
\end{equation}
where $\mathcal{H} = aH$, $Q = \Gamma \rho_m = - \dot{\Lambda}$, a prime means derivative with respect to conformal time, $\theta_{dm}$ is the dark matter {  velocity potential}, and $\phi$ is the gravitational potential.
In the limit of small scales, $k \gg \mathcal{H}$, these equations are reduced to
\begin{equation}
\theta'_{dm}+\mathcal{H}\theta_{dm}-k^2\Phi=0,
\end{equation}
\begin{equation}
\delta'_{dm}+\theta_{dm}=-\frac{aQ}{\rho_{dm}} \delta_{dm},
\end{equation}
\begin{equation}
-k^2 \Phi=\frac{a^2}{2} (\rho_{dm} \delta_{dm} + \rho_b \delta_b).
\end{equation}

\section{Data}
\label{Sec:Data}

In our analysis, we use the Joint Light-curve Analysis (JLA) supernovae data {  \cite{JLA}} together with the second release of Planck data \cite{Aghanim:2015xee} (hereafter TT+lowP), namely the high-$\ell$ Planck temperature data (in the range of $30< \ell <2508$) from the 100-, 143-, and 217- GHz half-mission TT cross-spectra, and the low-P data by the joint TT, EE, BB and TE likelihood (in the range of $2< \ell <29$).  It is worth mentioning that, compared to other recent SNe Ia compilations e.g. the Pantheon compilation \cite{Scolnic:2017caz}, the JLA sample has the advantage of allowing the light-curve recalibration with the model under consideration, which is an important issue when testing alternative cosmologies. We also consider current measurements of the local value of the Hubble-{Lema\^itre} parameter \cite{Riess:2019cxk} and  observations of  D/H abundance \cite{Cooke:2017cwo} as Gaussian priors on the $H_0$ and  $\Omega_{b0}h^2$ parameters. In the analysis which  $\mathcal A_l$ is left free to vary, we also consider CMB lensing data \cite{Ade:2015zua}. 

\section{Analysis and Results}
\label{Sec:Analysis}

We implemented the above sets of background and perturbation equations in the numerical Cosmic Linear Anisotropy Solving System (CLASS) code~\cite{Blas:2011rf} to generate the theoretical spectra of the model, while Monte Carlo Markov Chains (MCMC) {  analyses} for the  cosmological parameters constraints are obtained with Monte Python~\cite{Audren:2012wb} code.

In our analysis, we vary the usual cosmological parameters, namely, the physical baryon density, $\omega_b=\Omega_{b0}h^2$, the physical cold dark matter density, $\omega_{cdm}=\Omega_{dm0}h^2$, the optical depth, $\tau_{reio}$, the primordial scalar amplitude, $\mathcal A_s$, the primordial spectral index, $n_s$, the local Hubble-{Lema\^itre} parameter value $H_0$, in addition to the interaction parameter, $\alpha$. 
Also, we consider the lensing amplitude, $\mathcal A_l$, and the number of effective degrees of freedom at decoupling era, $N_{\text{eff}}$.
Indeed, it cannot be assumed a priori that interaction mechanisms allow for the same lensing scenario as expected from the standard cosmological model, i.e. $ \mathcal A_l = 1 $. Actually, as mentioned earlier, observational evidence seem to prefer large values of the lensing amplitude, which can be explained by allowing a closed curvature of the universe or interactions in the dark sector with dark matter production~\cite{heavens2, DiValentino:2019ffd}. 
At the same time, we also analyse an extension of the minimal model leaving $N_{\text{eff}}$ as a free parameter, in order to test how such a dark interaction can be degenerate with it.  In our analysis we vary the nuisance foreground parameters~\cite{Aghanim:2015xee} and consider purely adiabatic initial conditions. We choose to work with the Newtonian gauge, and we set the sum of neutrino masses fixed to $0.06$ eV. We work with flat priors for the cosmological parameters, also limiting the analysis to scalar perturbations with pivot scale $k_0=0.05$ $\rm{Mpc}^{-1}$. 

\begin{table*}[!t]
\centering
\caption{{
$68\%$ confidence limits for the cosmological parameters  using TT+lowP+JLA+ priors (Riess+Cooke). When the $\mathcal A_l$ parameter is considered as a free parameter in the analysis, we also add CMB lensing data.}
\label{tab:Tabel_LtCDM}}
\scalebox{0.7}{
\begin{tabular}{|c|c|c|c|c|c|c|c|c|}
\hline
\multicolumn{1}{|c|}{ }&
\multicolumn{2}{c|}{ $\Lambda$ CDM}& 
\multicolumn{2}{c|}{ $\Lambda$(t)CDM}&              
\multicolumn{2}{c|}{ $\Lambda$(t)CDM +  $\mathcal A_l$}& 
\multicolumn{2}{c|}{ $\Lambda$(t)CDM +  $\mathcal A_l$ + $N_{\text{eff}}$}\\ 
{Parameter}&
{\textbf{ mean}}& {\textbf{ best fit}}&
{\textbf{ mean}}& {\textbf{ best fit}}&
{\textbf{ mean}}& {\textbf{ best fit}}&
{\textbf{ mean}}& {\textbf{ best fit}}\\
\hline
$100\,\Omega_b h^2$ 	
& $2.245 \pm 0.018$ & 2.250
& $2.243 \pm 0.019$ & 2.249
& $2.260 \pm 0.021$ & 2.269
& $2.265 \pm 0.017$ & 2.272
\\
$\Omega_{c} h^2$	
& $0.1174 \pm 0.0019$ & 0.1185
& $0.1115 \pm 0.0079$ & 0.1094
& $0.1073 \pm 0.0100$ & 0.1063
& $0.1230 \pm 0.0088$ & 0.1245
\\
$\tau$
& $0.087 \pm 0.017$ & 0.104
& $0.088 \pm 0.018$ & 0.089
& $0.079 \pm 0.015$ & 0.088
& $0.074 \pm 0.018$ & 0.063
\\
$\ln 10^{10}A_s$  \footnotemark[1]
\footnotetext[1]{$k_0 = 0.05\,\Mpc^{-1}$.}
& $3.102 \pm 0.032$  & 3.140
& $3.106 \pm 0.034$  & 3.104
& $3.087 \pm 0.027$  & 3.101
& $3.083 \pm 0.036$  & 3.057
\\
$n_{s}$
& $ 0.9709 \pm 0.0056$ & 0.9679
& $ 0.9704 \pm 0.0057$ & 0.9692
& $ 0.9727 \pm 0.0059$ & 0.9735
& $ 0.9782 \pm 0.0062$ & 0.9851
\\
$H_0$
& $ 68.43 \pm 0.82 $ & 68.10
& $ 69.12 \pm 1.14 $ & 69.60
& $ 69.95 \pm 1.27 $ & 70.40
& $ 69.57 \pm 1.14 $ & 70.25
\\
$\alpha$
& - & -
& $ 0.037 \pm 0.050 $ & 0.045
& $ 0.059 \pm 0.066 $ & 0.060 
& $-0.018 \pm 0.047 $ & -0.030
\\
 $\mathcal A_l$
& - & -
& - & -
& $ 1.18 \pm 0.07 $ & 1.14
& $ 1.14 \pm 0.07 $ & 1.23
\\
$N_{\text{eff}}$
& - & -
& - & -
& - & -
& $ 3.22 \pm 0.14 $ & 3.30
\\
$\sigma_8$
& $ 0.827 \pm 0.014 $ & 0.845
& $ 0.841 \pm 0.024 $ & 0.841
& $ 0.839 \pm 0.026 $ & 0.849
& $ 0.817 \pm 0.027 $ & 0.798
\\
$\chi^2/2$
& $ - $ & 5977.0
& $ - $ & 5976.4
& $ - $ & 5977.8
& $ - $ & 5973.6
\\
\hline
\end{tabular}}
\end{table*} 
\begin{figure}
\centerline{\includegraphics[scale=0.4]{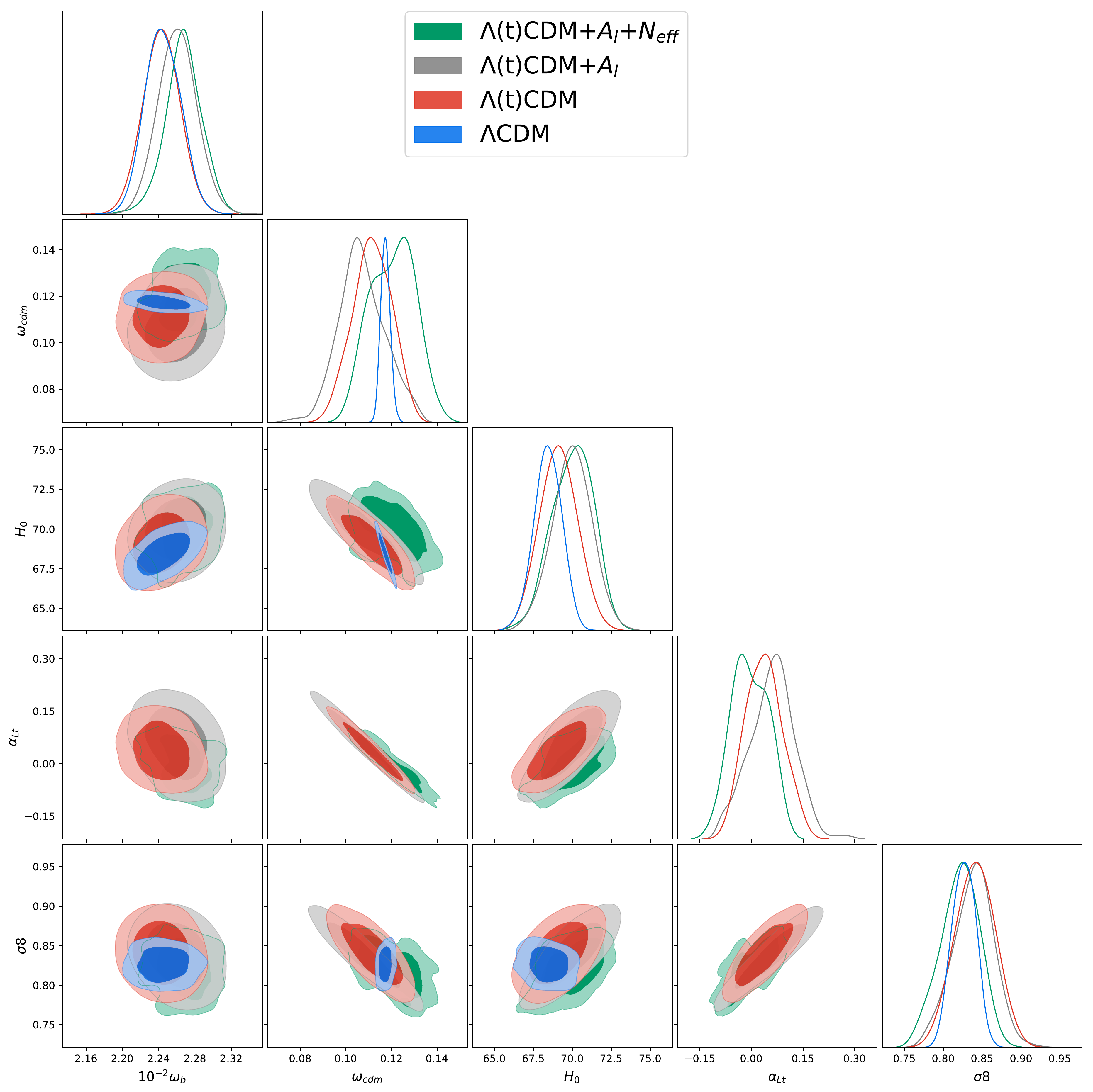} \hspace{.4in}}
\caption{Comparison between the constraints on the $\Lambda$CDM and $\Lambda$(t)CDM models using TT+lowP+JLA+ priors (Riess+Cooke). The confidence intervals are shown in Table \ref{tab:Tabel_LtCDM}.}
\label{fig:triangle_analysis}
\end{figure}

Our main results are {  shown} in Table {\ref{tab:Tabel_LtCDM}}, where the derived constraints on the cosmological parameters are displayed. In order to have a clearer comparison, we also show in Fig. \ref{fig:triangle_analysis} the corresponding confidence intervals at $68\%$ and $95\%$ and the posterior probability distribution for the most interesting behaviours. For instance, when the effective number of relativistic species is fixed in the standard value, $N_{\text{eff}} = 3.04$, a slightly positive mean value is preferred for the interaction parameter. This is in agreement with CMB joint analyses performed with different datasets \cite{Aurich:2017lck}, highlighting a contrast with the preference for negative values found with background and LSS probes \cite{Pigozzo,Ferreira:2017yby}. 

The value of the Hubble-{Lema\^itre} parameter is still in good agreement with the value constrained by the standard model using CMB data only, whereas for the $\Lambda$(t)CDM model and its extensions both the mean value and the best fit are slightly shifted to higher values, in better agreement with local measurements \cite{Riess:2019cxk}. For instance, for the combination of data sets used in our analysis, the discrepancy between the $\Lambda$CDM prediction and the value obtained from local distance estimators is $\simeq 3.41\sigma$ while for the $\Lambda$(t)CDM model one finds $\simeq 2.65\sigma$. The best concordance is obtained for the model $\Lambda$(t)CDM +  $\mathcal A_l$, i.e., $\simeq 2.25\sigma$. Besides, the positive correlation between the Hubble-{Lema\^itre} parameter and the interaction parameter, $\alpha$, seems to indicate a possibility to relax the $H_0$ tension even more at the cost of assuming an energy flux from matter to dark energy ($\alpha >0$). 

\begin{figure}
\centerline{\includegraphics[scale=1]{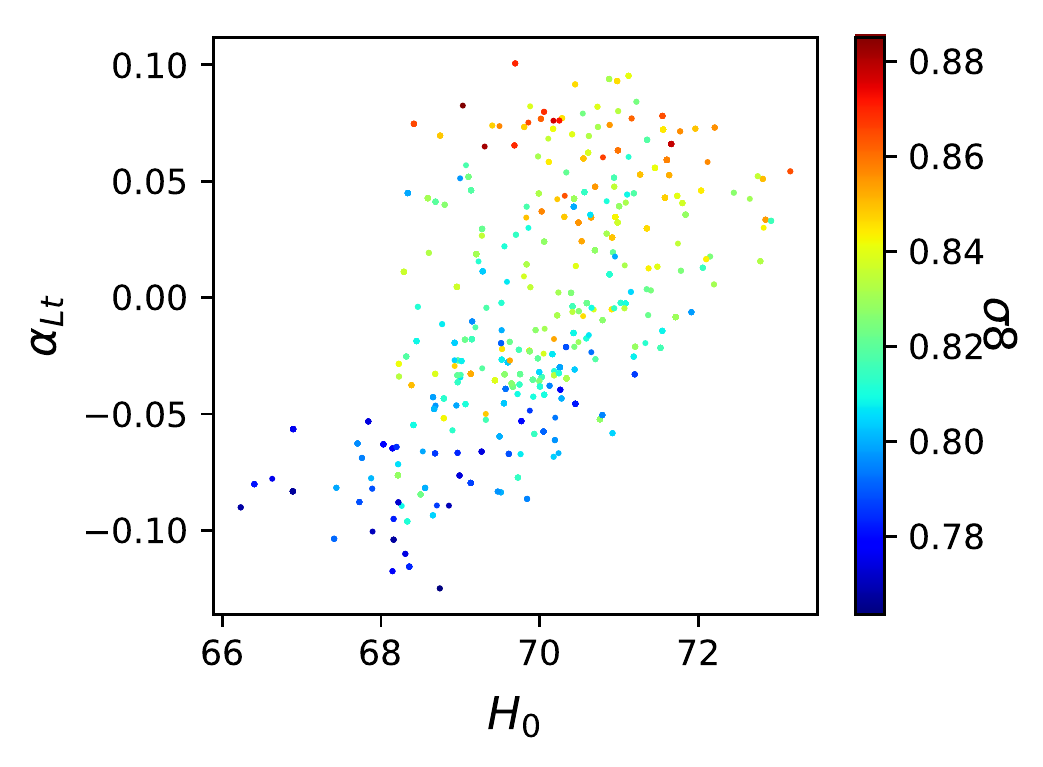} \hspace{.4in}}
\caption{Samples in the $H_0$ - $\alpha$ plane for the $\Lambda$CDM + $\mathcal A_l$ + $N_{\text{eff}}$ model using TT+lowP+JLA+ priors (Riess+Cooke), coloured by the value of the $\sigma_8$ parameter. The confidence intervals are given in the last column of Table \ref{tab:Tabel_LtCDM}.}
\label{fig:3D_H0_alphaLt_sigma8}
\end{figure}

When $N_{\text{eff}}$ is a free parameter of the model, the allowed interval of values shows a good agreement with previous $\Lambda$CDM analyses \cite{chineses,pedro}.  An important aspect worth mentioning is that the $\Lambda$(t)CDM + $\mathcal A_l$ + $N_{\text{eff}}$ model allows at the same time for lower values of $\sigma_8$ and slightly higher values of $H_0$ with respect to the standard model and, therefore, offers a possibility to alleviate the $H_0$ - $\sigma_8$ tension discussed earlier (see eg. \cite{Benetti:2017juy}). This behaviour is better visualised in Fig. \ref{fig:3D_H0_alphaLt_sigma8}, where the plane $H_0$ - $\alpha$ is { shown} with coloured $\sigma_8$ parameter values. Note that for negative values of $\alpha$, values of $H_0 >70$ $\rm{km/s/Mpc}$ and $\sigma_8<0.82$ are allowed at 1$\sigma$ level. 

\section{Final Remarks}
\label{Sec:Conclusions}

In this work we have considered a generalised Chaplygin gas model which provides a prime example for the class of unified dark { sector} models through the interacting parameter $\alpha$. While the standard $\Lambda$CDM model is recovered for $\alpha=0$, positive (or negative) values of such an interaction parameter indicate a creation of dark energy (or dark matter) by the dark sector.

Our results show that, although the $\alpha$ parameter value is always compatible with zero at 1$\sigma$, there is a preference for positive values when the CMB data are used. This is contrary to the results of previous works, where negative values of $\alpha$ were preferred at 2$\sigma$ level using LSS (2dFGRS), SNe Ia (JLA) and the position of the first peak of CMB \cite{Pigozzo}. At the same time, our results are in good agreement with more recent results where values of the interaction parameter close to zero were obtained from LSS and CMB data \cite{pedro,wiliam, Aurich:2017lck}.

We have also shown (see Table \ref{tab:Tabel_LtCDM}) that  negative values of $\alpha$ are obtained at the cost of a higher number of relativistic degrees of freedom (with respect to the standard $N_{\text{eff}} =3.046$). In this case, we note an interesting behaviour in the $H_0 - \sigma_8$ plane, since  lower values of the amplitude  of matter density fluctuations are allowed for higher values of $H_0$, relaxing both the $H_0$ and $\sigma_8$ tensions. It is important to emphasise that this is achieved not by breaking degeneracy  between the two parameters but by relaxing the constraint on them, i.e., at 1$\sigma$, the error on the two parameters is about twice what has been obtained  in the context of the standard cosmological model using the same dataset. Finally, the $\Lambda$(t)CDM model and its extensions, $\Lambda$(t)CDM + $\mathcal A_l$ and $\Lambda$(t)CDM + $\mathcal A_l$ + $N_{\text{eff}}$, reduce the discrepancy between the local and CMB $H_0$ measurements to $2.65\sigma$,  $2.25\sigma$ and $2.45\sigma$, respectively.

\section*{Acknowledgements}

MB thanks support of the Funda\c{c}\~{a}o Carlos Chagas Filho de Amparo \`{a} Pesquisa do Estado do Rio de Janeiro (FAPERJ - fellowship {\textit{Nota 10}}), and Istituto Nazionale di Fisica Nucleare (INFN), sezione di Napoli, iniziative specifiche QGSKY. SC is supported by CNPq (Brazil) with grant no. 307467/2017-1. JSA acknowledges support from CNPq (grants no. 310790/2014-0 and 400471/2014-0) and FAPERJ (grant no. 204282). The authors thank the use of CLASS and Monte Python codes. We also  acknowledge the use of  the High Performance Computing Centre at the Universidade Federal do Rio Grande do Norte (NPAD/UFRN) and the Observat\'orio Nacional (ON) computational support.

\end{document}